\documentclass[conference]{IEEEtran}
\IEEEoverridecommandlockouts
\usepackage{cite}
\usepackage{textcomp}
\usepackage{xcolor}
\usepackage{times}
\usepackage{helvet}
\usepackage{courier}
\usepackage[hyphens]{url}
\usepackage{graphicx}
\usepackage{caption}
\usepackage{amsmath}
\usepackage{algpseudocode}
\usepackage{algorithm}
\usepackage{array}
\usepackage{colortbl}
\usepackage{hhline}
\usepackage[T1]{fontenc}
\usepackage{rotating}
\usepackage{microtype}
\usepackage{multirow}
\usepackage[utf8]{inputenc}
\usepackage{url}
\usepackage{booktabs}
\usepackage{amsfonts}
\usepackage{nicefrac}
\usepackage{latexsym}
\usepackage{booktabs}
\usepackage{tabularx}
\usepackage{subcaption}
\usepackage{inconsolata}
\usepackage{hyperref}

\def\BibTeX{{\rm B\kern-.05em{\sc i\kern-.025em b}\kern-.08em
    T\kern-.1667em\lower.7ex\hbox{E}\kern-.125emX}}

\begin{document}
\sloppy

\title{Gibberish is All You Need for Membership Inference Detection in Contrastive Language-Audio Pretraining}

\author{
    \IEEEauthorblockN{Ruoxi Cheng$^{1*}$, Shuirong Cao$^{2*}$, Yizhong Ding$^{1*}$, Shitong Shao$^{3}$, Zhiqiang Wang$^{1*\dag}$}
    \IEEEauthorblockA{
        \textit{Beijing Electronic Science and Technology Institute, Beijing, China} \\
        \textit{AVIC Nanjing
Engineering Institute of Aircraft Systems, Nanjing, China} \\
        \textit{Hong Kong University of Science and Technology (Guangzhou), Guangzhou, China} \\
        }
    \thanks{$^*$ Contributed equally to this work. $^\dag$ Corresponding to \href{mailto:wangzq@besti.edu.cn}{wangzq@besti.edu.cn}. Supported by the Fundamental Research Funds for the Central Universities (Grant Number: 3282024050).}
}

\maketitle

\begin{abstract}
Audio can disclose PII, particularly when combined with related text data. Therefore, it is essential to develop tools to detect privacy leakage in Contrastive Language-Audio Pretraining(CLAP). Existing MIAs need audio as input, risking exposure of voiceprint and requiring costly shadow models. We first propose PRMID, a membership inference detector based probability ranking given by CLAP, which does not require training shadow models but still requires both audio and text of the individual as input. To address these limitations, we then propose USMID, a textual unimodal speaker-level membership inference detector, querying the target model using only text data. We randomly generate textual gibberish that are clearly not in training dataset. Then we extract feature vectors from these texts using the CLAP model and train a set of anomaly detectors on them. During inference, the feature vector of each test text is input into the anomaly detector to determine if the speaker is in the training set (anomalous) or not (normal). If available, USMID can further enhance detection by integrating real audio of the tested speaker. Extensive experiments on various CLAP model architectures and datasets demonstrate that USMID outperforms baseline methods using only text data.
\end{abstract}

\begin{IEEEkeywords}
membership inference, CLAP models, privacy
\end{IEEEkeywords}

\section{Introduction}
Microphones in Internet of Things (IoT) devices~\cite{abdul2015internet} like phones can lead to unintended inferences from audio~\cite{shah2021evaluating,feng2022user,10095737,li2023membership}. Vocal features and linguistic content can reveal personally identifiable information (PII)~\cite{schwartz2011pii} like biometric identity and socioeconomic status. Combining audio with text data increases susceptibility to inference attacks. Thus, developing tools to detect privacy leakage in text-audio models like contrastive language-audio pre-training(CLAP)~\cite{elizalde2023clap,zhao2023clap,10095969} is essential.

Traditional methods like membership inference attacks (MIAs)~\cite{shokri2017membership} focus on determining whether a specific data sample was used for model training. Research on MIAs for multimodal contrastive learning (MCL)~\cite{yuan2021multimodal} like Contrastive Language-Image Pretraining(CLIP)~\cite{radford2021learning} is extensive~\cite{ko2023,li2024identity,hintersdorf2024does}, but little attention is given to CLAP.

Traditional MIAs train shadow models to simulate target model’s behavior~\cite{abdullah2021hear, chen2023slmia, tseng2021membership}, which requires high computational costs, particularly for multimodal models like CLAP. We first propose PRMID, which uses the probability ranking provided by CLAP for membership inference detection, thereby avoiding the computational costs of shadow models.

\begin{figure}[t]
  \centering
  \includegraphics[width=0.4\textwidth]{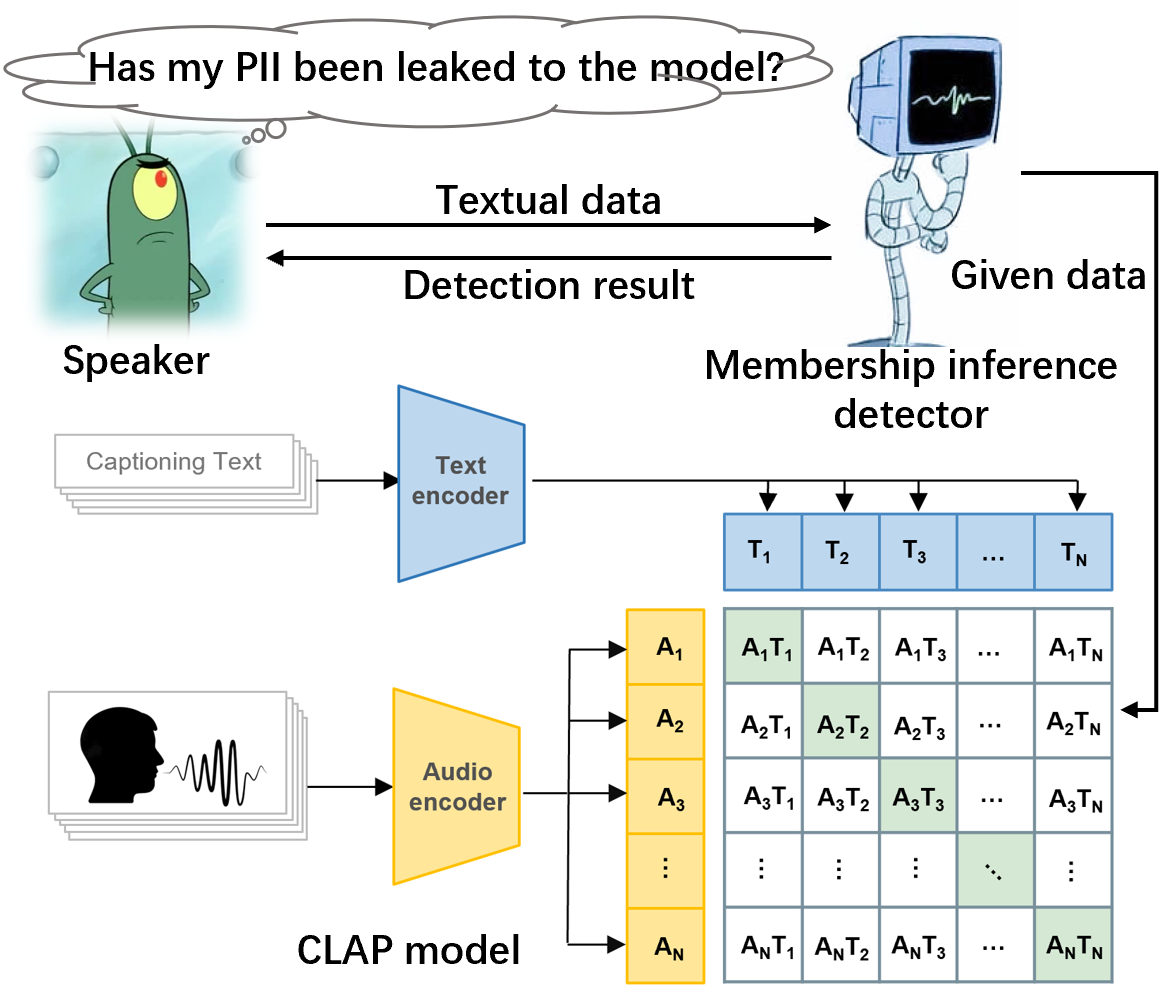}
  \caption{\small Current MIAs on MCL always query with dual-modal data of the tested individual for inference, while our goal is to avoid this.}
  \label{fig:example}
\end{figure}

However, current MIAs for MCL as well as PRMID often rely on dual-modal data inputs~\cite{hu2022m}, which may lead to new leakage, as one modal of the pair might not have been exposed to the risky target model. Therefore, a detector that does not query CLAP with explicitly matched audio-text pair of speaker (see an example in Figure~\ref{fig:example}) is desirable. This concept is known as multimodal data protection~\cite{liu2024multimodal}. 

To address these limitations, we propose USMID, a textual unimodal speaker-level membership inference~\cite{miao2022no} detector for CLAP models, which queries the target model with only text data. Specifically, we introduce a feature extractor that maps text data to feature vectors through CLAP-guided audio optimization. We then generate sufficient text gibberish that clearly does not match any text description in training dataset.

 \begin{figure}[t]
    \centering
    \begin{minipage}{0.4\textwidth}
        \includegraphics[width=\linewidth]{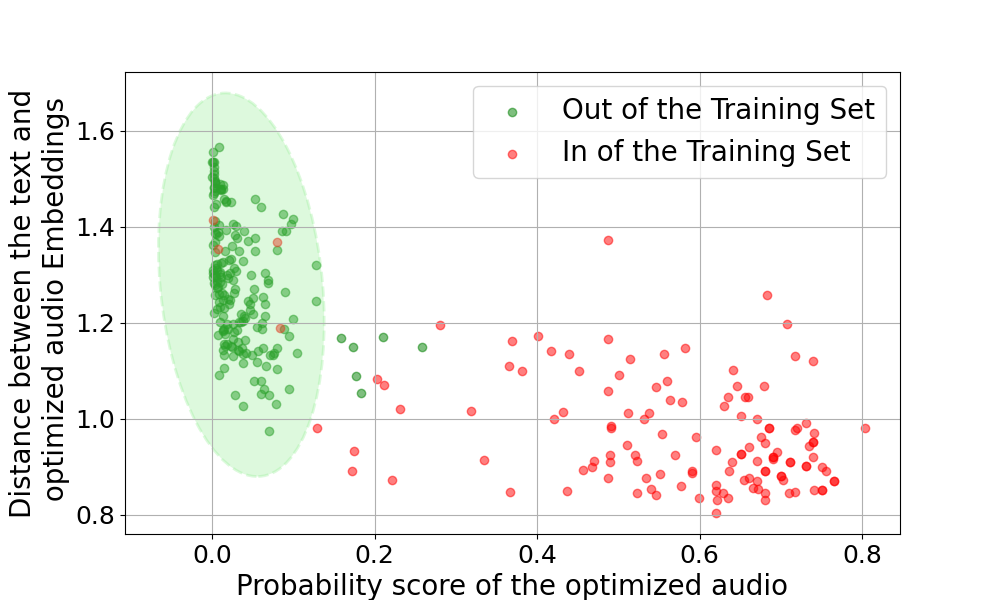}
    \end{minipage}\hfill
    \begin{minipage}{0.48\textwidth}
        \caption{\small Optimization of audio is guided by a CLAP model trained on LibriSpeech dataset where each person has 50 audios. Distance between the embeddings of optimized audio and tested text, and probability score of the tested text among gibberish, can clearly distinguish between samples within and outside the training set of target CLAP model.}
        \label{fig:separation}
    \end{minipage}
\end{figure}

As shown in Figure~\ref{fig:separation}, we observe a distinct separation between the features of gibberish and members in training set.

Based on this observation, we train multiple anomaly detectors using the feature vectors of generated text gibberish, creating an anomaly detection voting system. During testing, USMID inputs the feature vectors of test text into the voting system to determine if the corresponding speaker is in(anomalous) or out(normal) of the training set. 

Our contributions are summarized as follows:
\begin{itemize}
\item We are the first to study membership inference detection in CLAP, constructing several audio-text pair datasets and trained various architectures of CLAP models.
\item We introduce USMID, the first speaker-level membership inference detector for CLAP, which avoids exposing audio data to risky target model and the high cost for training shadow models in traditional MIAs.
\item Extensive experiments show that USMID outperforms all baselines even using only text PII for query.
\end{itemize}

\section{Related Work}

\subsection{Contrastive Language-Audio Pretraining}

Contrastive language-audio pretraining has significantly improved multimodal representation learning~\cite{wu2023large}. Techniques like DSCLAP and T-CLAP enhance domain-specific applications and temporal alignment, showcasing the effectiveness of integrating language and audio~\cite{li2024advancing}.

\subsection{Membership Inference in Automatic Speech Recognition}

Recent studies show that automatic speech recognition (ASR) systems are vulnerable to MIAs\cite{li2023membership,shah2021evaluating}.These MIAs typically rely on costly shadow models\cite{chen2023slmia,miao2022no} and require real audio as input to target model\cite{abdullah2021hear}, which may lead to new leakage.

\section{Threat Model}

Consider a CLAP model \(M\) trained on a dataset \(D_{\text{train}}\). Each sample \(s_i = (t_i, x_i)\) in \(D_{\text{train}}\) contains the PII of a speaker, consisting of a textual description \(t_i\) and its corresponding audio \(x_i\). For distinct indices \(i \neq j\), it is possible for \(t_i = t_j\) while \(x_i \neq x_j\), indicating that multiple non-identical audio samples may exist for the same speaker.

\textbf{Detector's Goal.} The detector aims to probe potential leakage of a speaker's PII through the target CLAP model \(M\), seeking to determine whether any PII of the speaker were included in the training set \(D_{\text{train}}\).
For a speaker with textual description \(t\), the detector aims to determine whether there exists a PII sample \((t_i, x_i) \in D_{\text{train}}\) such that \(t_i = t\).

Note that our goal is not to detect a specific text-audio pair \((t, x)\), but rather to identify the existence of any pair with textual description \(t\). This is because that multiple audio samples of the same speaker may be used for training, any of which could contribute to potential PII leakage.

\textbf{Detector’s Knowledge and Capability.} The detector operates with black-box access to \(M\), i.e., it can query \(M\) and observe the outputs, but does not know the model architecture of \(M\), the parameter values, or the training algorithms. For the target textual description $t$, depending on the application scenarios, the detector may or may not have actual audios corresponding to $t$. However, if the detector does have the corresponding audio samples, it cannot include them in its queries to \(M\) due to privacy concerns. Additionally, the detector is unable to modify \(M\) or access its internal state.

\begin{figure}[t]
  \centering
  \includegraphics[width=0.48\textwidth]{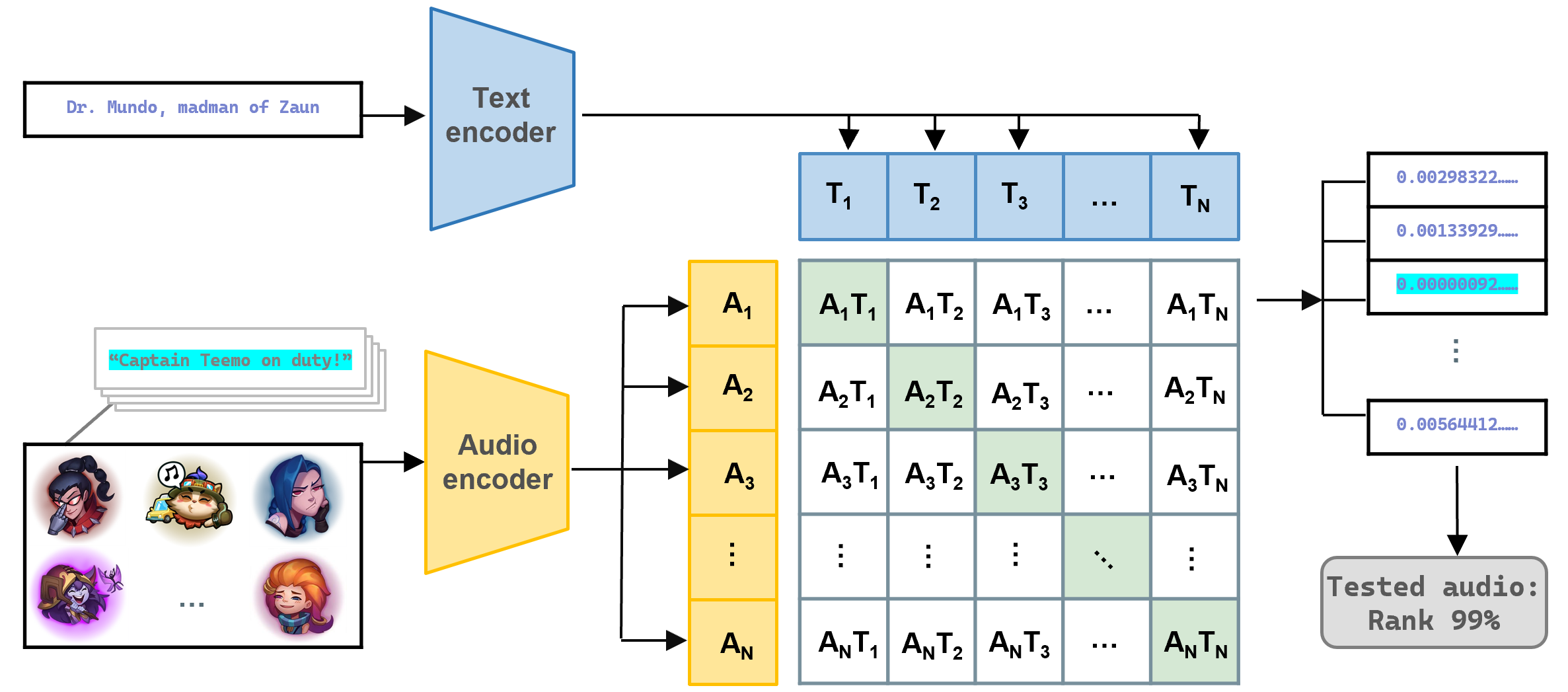}
  \caption{\small To determine whether a person's text is in the training set, we input his audio alongside a collection of other individuals' audios into the CLAP model. The model then generates a probability distribution based on the matching scores, which we use to conduct inference.}
  \label{fig:testaudio}
\end{figure}

\begin{figure}[t] \centering \includegraphics[width=0.48\textwidth]{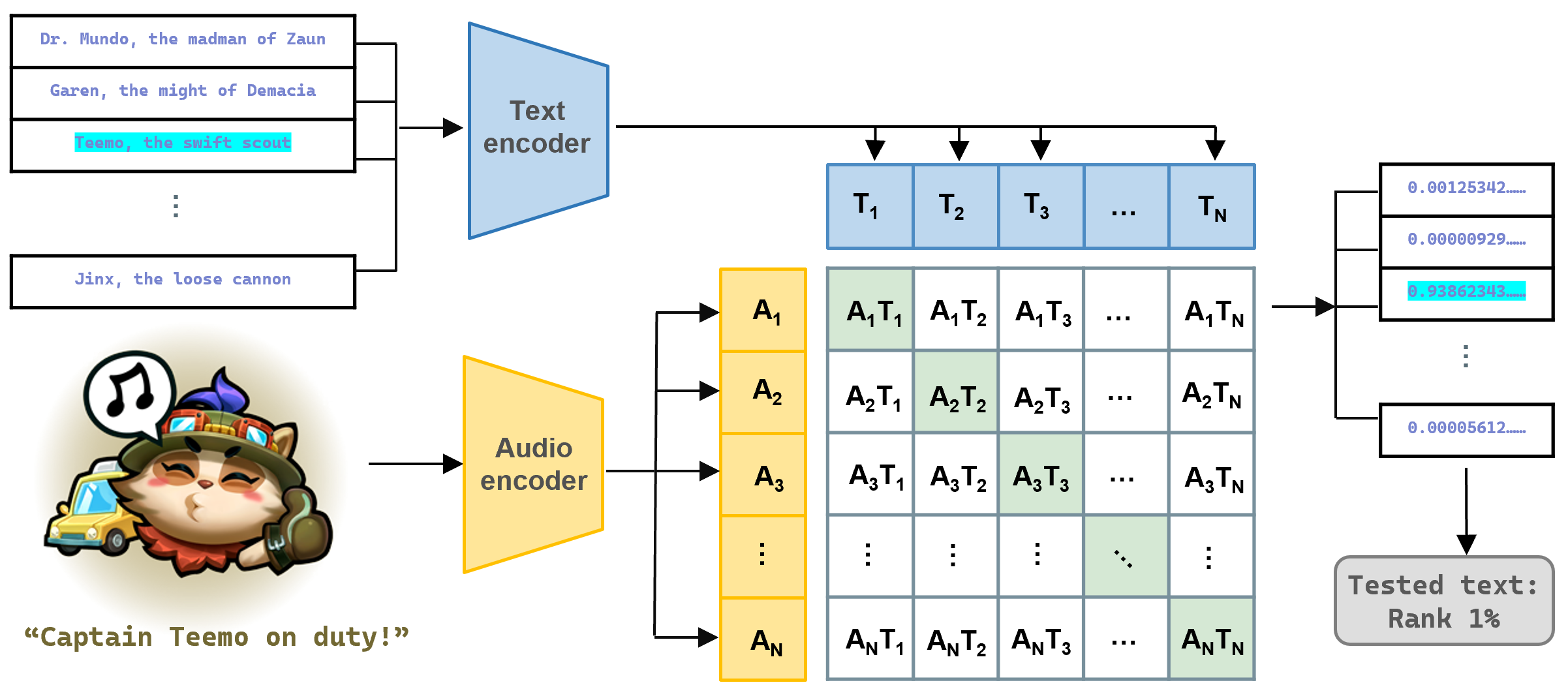} \caption{\small To determine whether a person's audio is in the training set, we input his text alongside a collection of texts from other individuals.} \label{fig
} \end{figure}

\section{Methodology}

\begin{figure*}[t]
  \centering
  \includegraphics[width=0.9 \textwidth]{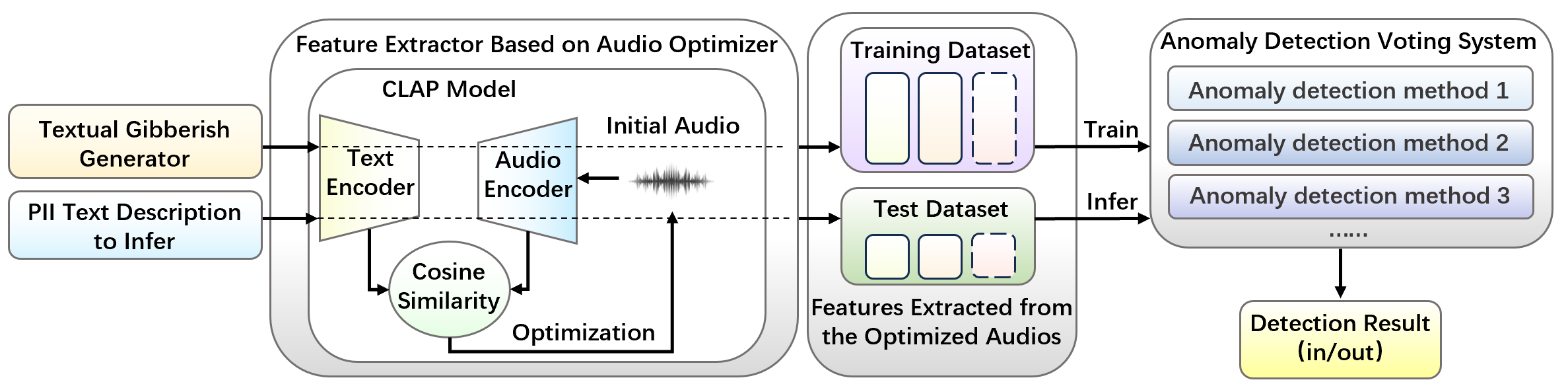}
  \caption{\small Overview of USMID.}
  \label{fig:USMID}
\end{figure*}

\subsection{Probability Ranking Membership Inference Detector}

CLAP is trained to maximize cosine similarity between audio and text features of members. Thus, if one modality of a member is provided to target model, the corresponding other modality data typically yields a higher probability score in the calculated distribution when input alongside other samples.

Based on this, we propose PRMID (Probability Ranking Membership Inference Detector) as shown in Figure~\ref{fig:testaudio}.

\textbf{Probability Distribution Evaluated by CLAP.} We first match the tested audio \( x \) with tested text \( t \) and a set of textual gibberish \( \mathcal{G} = \{g_1, g_2, \ldots, g_\ell\} \). We use CLAP to obtain the probability distribution \( \mathcal{P} = \{P(t), P(g_1), P(g_2), \ldots, P(g_\ell)\} \), where \( P(t) + P(g_1) + P(g_2) + \ldots + P(g_\ell) = 1 \).

\textbf{Membership Inference through Ranking.} We define the rank of the tested text \( t \) within the probability distribution \( \mathcal{P} \) as \( r_t = P(t) \). We conduct \( N \) repeated experiments, generating \( \ell \) gibberish samples in each trial. Each experiment yields a probability distribution \( \mathcal{P} \), which enables us to analyze \( r_t \).

We set thresholds \( T_1 \) and \( T_2 \) for top \( k \%\) and bottom \( k \%\), where \( k \%\) is a specified percentage (for example, 1\%). 

We consider three scenarios below:

\begin{itemize}
    \item If count of \( r_t \) in top \( k \% \) exceeds \( T_1 \) across \( N \) experiments, we infer that both \( t \) and \( x \) are present in \( D_{\text{train}} \).
    
    \item If count of \( r_t \) in bottom \( k \% \) exceeds \( T_2 \) across \( N \) experiments, \( t \) is outside of \( D_{\text{train}} \), while \( x \) remains within.
    
    \item A sample is classified as random if \( r_t \) exhibits a uniform distribution across all \( \ell + 1 \) options. Specifically, the expected probability for any rank is \( \frac{1}{\ell + 1} \). If the observed frequencies for each rank fall within the expected range of \( \frac{N}{\ell + 1} \), we conclude that \( t \) is outside of \( D_{\text{train}} \), with the status of \( x \) remaining undetermined.
\end{itemize}

\textbf{Membership inference for Audio.} In reverse inference, we can swap the roles of audio and text and repeat the inference process above as illustrated in Figure~\ref{fig
}, allowing membership inference for both modalities.

\subsection{Unimodal Speaker-Level Membership Inference Detector}

While PRMID requires both audio and text inputs from the individual as input for the target model, this can introduce new privacy risks, as the target model may not have previously encountered dual-modal PII of that individual.

To address this limitation, we propose USMID (unimodal detector for membership inference detection). This detector is designed to ascertain whether the PII of a speaker is included in the training set of target CLAP model \( M \), under the condition that only the speaker's textual description is provided to \( M \).

An overview of USMID is illustrated in Figure~\ref{fig:USMID}. Firstly, for a textual description $t$, we develop a feature extractor to map $t$ to a feature vector, through audio optimization guided by CLAP. Then, we make the key observation that \emph{textual gibberish like ``dv3*4l-XT0''—random combinations of numbers and symbols clearly do not match any textual descriptions in training set}, and hence the detector can generate large amount of textual gibberish that are known out of $D_{\text{train}}$. Using feature vectors extracted from these gibberish, detector can train multiple anomaly detectors to form an anomaly detection voting system. Finally, during inference phase, the features of the target textual description are fed into the system, and the inference result is determined through voting. Furthermore, when actual audio samples corresponding to the textual description are available, the detector can leverage them to perform clustering on feature vectors of the test samples to enhance detection performance.

\textbf{Feature Extraction through CLAP-guided Audio Optimization.} The feature extraction for a textual description $t$ involves iterative optimization of an audio $x$, to maximize the correlation between the embeddings of $t$ and $x$ produced by the target CLAP model. The extraction process, described in Algorithm~\ref{alg:feature-extraction}, 
iterates for $n$ epochs; and within each epoch, an audio is optimized for $m$ iterations, to maximize the cosine similarity between its embedding of CLAP and that of target textual description. The average optimized cosine similarity $S$ and standard deviation of optimized audio embeddings $D$ are extracted as the features of $t$ from model $M$.

\begin{algorithm}[tb]
\caption{CLAP-guided Feature Extraction}
\label{alg:feature-extraction}
\textbf{Input}: Target CLAP model \( M \), textual description \( t \) \\
\textbf{Output}: Mean optimized cosine similarity \( S \), standard deviation of optimized audio embeddings \( D \)
\begin{algorithmic}[1]
\State $n \gets$ number of epochs
\State $m \gets$ number of optimization iterations per epoch
\State $\mathcal{S} \gets \emptyset$, $\mathcal{V} \gets \emptyset$
\State $v_t \gets M(t)$ \Comment{Obtain text embedding from $M$}
\For{$i = 1$ \textbf{to} $n$}
    \State $x_0 \gets \textup{Rand}()$ \Comment{Randomly generate an initial audio}
    \For{$j = 0$ \textbf{to} $m-1$}
        \State $v_{x_j} \gets M(x_j)$ \Comment{Obtain audio embedding from $M$}
        \State $x_{j+1} \gets \arg \max_{x_j} \frac{v_t \cdot v_{x_j}}{\|v_t\| \, \|v_{x_j}\|}$ \Comment{Update audio to maximize cosine similarity}
    \EndFor
    \State $S_i \gets \frac{v_t \cdot v_{x_m}}{\|v_t\| \, \|v_{x_m}\|}$ \Comment{Optimized similarity for epoch $i$}
    \State $\mathcal{S} \gets \mathcal{S} \cup \{S_i\}$, $\mathcal{V} \gets \mathcal{V} \cup \{v_{x_m}\}$
\EndFor
\State $S \gets \frac{1}{n} \sum_{S_i \in \mathcal{S}} S_i$
\State $\bar{v} \gets \frac{1}{n} \sum_{v \in \mathcal{V}} v$
\State $D \gets \sqrt{\frac{1}{n} \sum_{v \in \mathcal{V}} \|v - \bar{v}\|^2}$
\State \textbf{return} $S$, $D$
\end{algorithmic}
\end{algorithm}

\textbf{Generation of Textual Gibberish.} USMID starts the detection process with generating a set of $\ell$ gibberish strings $\mathcal{G} = \{g_1,g_2,\ldots,g_{\ell}\}$, which are random combinations of digits and symbols with certain length. As these gibberish texts are randomly generated at the inference time, with overwhelming probability that they did not appear in the training set. Applying the proposed feature extraction algorithm on ${\cal G}$, we obtain $\ell$ feature vectors ${\cal F} = \{f_1,f_2,\ldots,f_{\ell}\}$ of the gibberish texts.

\textbf{Training Anomaly Detectors.} Motivated by the observations in Figure3 that feature vectors of the texts in and out of the training set of $M$ are well separated, we propose to train an anomaly detector using ${\cal F}$, such that texts out of $D_{\text{train}}$ are considered ``normal'', and the problem of membership inference on $t$ is converted to anomaly detection on its feature vector. More specifically, $t$ is classified as part of $D_{\text{train}}$, if its feature vector is detected ``abnormal'' by the trained anomaly detector. Specifically in USMID, we train several anomaly detection models on ${\cal F}$, such as Isolation Forest~\cite{liu2008isolation}, LocalOutlierFactor~\cite{cheng2019outlier} and AutoEncoder~\cite{chandola2009anomaly}. These models constitute an anomaly detection voting system that will be used for membership inference on the test textual descriptions.

\textbf{Textual Membership Inference through Voting.} For each textual description $t$ in the test set, USMID first extracts its feature vector $f$ using Algorithm~\ref{alg:feature-extraction}, and then feeds $f$ to each of the obtained anomaly detectors to cast a vote on whether $t$ is an anomaly. When the total number of votes exceeds a predefined detetion threshold $N$, $t$ is determined as an anomaly, i.e., PII with textual description $t$ is used to train the CLAP model $M$; otherwise, $t$ is considered normal and no PII with $t$ is leaked through training of $M$.

\textbf{Enhancement with Real audios.} At inference time, if real audios of the test texts are available at the detector (e.g., audios of a person), they can be used to extract an additional feature measuring the average distance between the embeddings of real audios and those of optimized audios using the CLAP model, using which the feature vectors of the test texts can be clustered into two partitions with one in $D_{\text{train}}$ and another one out of $D_{\text{train}}$. This adds an additional vote for each test text to the above described anomaly detection voting system, potentially facilitating the detection accuracy.

Specifically, for each test text $t$, the detector is equipped with a set of $c$ real audios $\{x_{\mathrm{real}}^1, x_{\mathrm{real}}^2, \ldots, x_{\mathrm{real}}^c\}$. Similar to the feature extraction process in Algorithm \ref{alg:feature-extraction}, over $k$ epochs with independent initializations, $k$ optimized audios $\{x_{\mathrm{opt}}^1, x_{\mathrm{opt}}^2, \ldots, x_{\mathrm{opt}}^k\}$ for $t$ are obtained under the guidance of the CLAP model. Then, we apply a pretrained feature extraction model $F$ (e.g.,DeepFace for face audios) to the real and optimized audios to obtain real embeddings $\{v_{\mathrm{real}}^1, v_{\mathrm{real}}^2, \ldots, v_{\mathrm{real}}^c\}$ and optimized embeddings $\{v_{\mathrm{opt}}^1, v_{\mathrm{opt}}^2, \ldots, v_{\mathrm{opt}}^k\}$. Finally, we compute the average pair-wise $\ell_2$ distance between the real and optimized embeddings, denoted by $R$, over $c \cdot k$ pairs, and use $R$ as an additional feature of the text $t$.

For a batch of $B$ test texts $(t_1, t_2, \ldots, t_B)$, we extract their features $((S_1, D_1, R_1), (S_2, D_2, R_2), \ldots, (S_B, D_B, R_B))$ first. Feeding the first two features $S_i$ and $D_i$ into a trained anomaly detection system, each text $t_i$ obtains an anomaly score based on the number of detectors that classify it as abnormal. Additionally, the $K$-means algorithm with $K=2$ partitions the feature vectors $\{(S_i, D_i, R_i)\}_{i=1}^B$ into ``normal'' cluster and an ``abnormal'' clusters, contributing another vote to the anomaly score of each instance. Finally, membership inference is performed by comparing the total votes received to a detection threshold $N'$.

\section{Evaluations}
\label{sec:evalu}
We evaluate the performance of USMID, for speaker-level membership inference using only text PII of the individual.

\textbf{Dataset Construction.} In addition to LibriSpeech~\cite{7178964}, we built a speaker recognition dataset based on CommonVoice18.0~\cite{DBLP:journals/corr/abs-1912-06670}, which covers various social groups and has richer background information. Specifically, 3,000 speakers (1,500 for training and 1,500 for verification) were selected from CommonVoice, and their audio files were accompanied by unique user PII like ID, age, gender, and region information; then for each user ID, we used GPT-4o to generate detailed background description based on their PII; finally, these expanded background descriptions and audio files corresponding to each user ID constituted the training set of CLAP.

By doing this, we obtained basic facts about who is in the training set and who is not. For each type of content, we created two datasets: one with 1 audio clip per person and another with 50 audio clips per person.

\textbf{Models.} In our CLAP model, audio encoder uses HTSAT\cite{chen2022hts}, which is transformer with 4 groups of swin-transformer blocks. We use the output of its penultimate layer (a 768-dimensional vector) as the output sent to the projection MLP layer. Text encoder uses RoBERTa, which converts input text into a 768-dimensional feature vector. We apply a 2-layer MLP with ReLU activation \cite{agarap2018deep} to map the audio and text outputs to 512 dimensions for final representation.

\textbf{Evaluation Metrics.} USMID's effectiveness is assessed using Precision, Recall, and Accuracy metrics, measuring anomaly prediction accuracy, correct anomaly identification, and overall prediction correctness, respectively.

\textbf{Baselines.} Current speaker-level membership inference detection methods require detector to query target model with real audio. Most MIAs involve training shadow models, which can be particularly costly for multimodal LLMs. We empirically compare the performance of USMID with PRMID and the following SOTA inference methods. The audio encoders for Audio Auditor and SLMIA-SR are LSTM, for AuditMI they are Transformer, and for PRMID and USMID, they are CLAP.

\begin{itemize}
    \item {\bf Audio Auditor}~\cite{miao2022no} trains shadow models and extracts audio features for inference.
    \item {\bf SLMIA-SR}~\cite{chen2023slmia} employs a shadow speaker recognition system to train attack model.
    \item {\bf AuditMI}~\cite{teixeira2024exploring} trains shadow model using input utterances and features from model outputs.
\end{itemize}

All experiments are performed using four NVIDIA GeForce RTX 4090 GPUs. Each experiment is repeated for 10 times, and the average values and the standard deviations are reported.

\subsection{Results} 
On training anomaly detectors, we randomly generated $\ell = 100$ textual gibberish (some of them are shown in Table~\ref{table:gibberish}).

\begin{table}[t]
\centering
%\tiny
\caption{\small Samples of randomly generated gibberish.}
\label{table:gibberish}
\begin{tabular}{|c|c|c|}
\hline
\b+dhu!fh4Y35x9dew & 453fKer5(s=pnI<S & fewf3\_;\@fw/\\
\hline
d3|\%5f3et45tfG\textbackslash\_ & 4terbtyh<E\{43ter & 5gggtb-~=64hgF \\
\hline
\#43GRGgrt54Hdg & '2\_:gtrg6[45gerg3b & g*|<trgrtthg4t|3/ \\
\hline
\end{tabular}
\end{table}

\begin{table}[t]
\renewcommand{\arraystretch}{0.6} % Adjust row spacing
\tiny % Adjust font size to be smaller (1/3 of normal size)
\centering
\caption{\small Comparison with baseline methods.}
\resizebox{0.48\textwidth}{!}{
\begin{tabular}{@{}lccccc@{}}
\toprule
\textbf{Architecture} & \textbf{Number of Audios per} & \textbf{Method} & \textbf{Precision} & \textbf{Recall} & \textbf{Accuracy} \\
 & \textbf{person in training set} & & & & \\
\midrule
\multirow{13}{*}{LibriSpeech} & \multirow{4}{*}{1} 
& Audio Auditor & 63.38 ± 0.24 & 73.24 ± 0.33 & 65.19 ± 0.27 \\
 & & SLMIA-SR & 75.21 ± 0.18 & 88.64 ± 0.14 & 83.42 ± 0.21 \\
 & & AuditMI & 82.57 ± 0.21 & 95.26 ± 0.26 & 87.91 ± 0.24 \\
 & & PRMID & 85.32 ± 0.18 & 95.58 ± 0.22 & 89.75 ± 0.17 \\
 & & USMID & \textbf{86.49 ± 0.19} & \textbf{96.49 ± 0.23} & \textbf{91.27 ± 0.15} \\
\cmidrule{2-6}
 & \multirow{4}{*}{50} 
& Audio Auditor & 65.59 ± 0.23 & 80.13 ± 0.16 & 66.59 ± 0.29 \\
 & & SLMIA-SR & 76.19 ± 0.31 & 90.07 ± 0.18 & 84.33 ± 0.25 \\
 & & AuditMI & 83.41 ± 0.14 & 98.04 ± 0.09 & 88.16 ± 0.13 \\
 & & PRMID & 86.15 ± 0.16 & 95.87 ± 0.24 & 90.12 ± 0.19 \\
 & & USMID & \textbf{88.12 ± 0.26} & \textbf{98.76 ± 0.12} & \textbf{93.07 ± 0.16} \\
\midrule
\multirow{13}{*}{CommonVoice} & \multirow{4}{*}{1} 
& Audio Auditor & 54.85 ± 0.23 & 68.22 ± 0.19 & 60.52 ± 0.21 \\
 & & SLMIA-SR & 65.39 ± 0.36 & 76.91 ± 0.27 & 70.48 ± 0.24 \\
 & & AuditMI & 71.43 ± 0.28 & 81.45 ± 0.41 & 74.36 ± 0.18 \\
 & & PRMID & 72.35 ± 0.23 & 84.52 ± 0.20 & 78.43 ± 0.18 \\
 & & USMID & \textbf{74.96 ± 0.25} & \textbf{86.01 ± 0.22} & \textbf{81.79 ± 0.15} \\
\cmidrule{2-6}
 & \multirow{4}{*}{50} 
& Audio Auditor & 56.11 ± 0.33 & 73.58 ± 0.27 & 61.35 ± 0.25 \\
 & & SLMIA-SR & 66.28 ± 0.21 & 79.27 ± 0.34 & 72.18 ± 0.22 \\
 & & AuditMI & 73.52 ± 0.17 & 84.81 ± 0.28 & 75.64 ± 0.23 \\
 & & PRMID & 75.12 ± 0.19 & 88.26 ± 0.18 & 80.98 ± 0.14 \\
 & & USMID & \textbf{76.47 ± 0.12} & \textbf{89.46 ± 0.32} & \textbf{82.33 ± 0.19} \\
\bottomrule
\end{tabular}
}
\label{table:baseline}
\end{table}

\begin{table}[t]
\centering
\tiny
\renewcommand{\arraystretch}{0.6}
\caption{\small Comparison of training time, GPU memory consumption, and inference time per sample with baselines on LibriSpeech dataset.}
\resizebox{0.4\textwidth}{!}{  % 将表格宽度调整为页面的一半
\begin{tabular}{@{}lccc@{}}
\toprule
\textbf{Method} & \textbf{Train Time} & \textbf{GPU Memory} & \textbf{Inference Time} \\
\midrule
Audio Auditor & 7.5h & 11.3GB & 0.359s \\
SLMIA-SR & 9h & 13.7GB & 0.406s \\
AuditMI & 80h & 49.5GB & 2.375s \\
USMID & 3.7h & 24.3GB & 0.628s \\
\bottomrule
\end{tabular}
}
\label{table:costs}
\end{table}

\begin{table}[t]
\renewcommand{\arraystretch}{0.6} % Adjust row spacing
\tiny % Adjust font size to be smaller (1/3 of normal size)
\centering
\caption{\small Comparison of performance with a given audio.}
\resizebox{0.48\textwidth}{!}{
\begin{tabular}{@{}lcccccc@{}}
\toprule
\textbf{Architecture} & \multicolumn{1}{c}{\textbf{Number of audios per}} & \textbf{USMID} & \textbf{Precision} & \textbf{Recall} & \textbf{Accuracy} \\
& \multicolumn{1}{c}{\textbf{person in training set}} & & & &\\
\midrule
\multirow{4.5}{*}{LibriSpeech} & \multirow{2}{*}{1} 
& Text only & 86.49 ± 0.19  & 96.49 ± 0.23 & 91.27 ± 0.15  \\
 & & With 1 audio & \textbf{89.21 ± 0.14} & \textbf{98.68 ± 0.18} & \textbf{93.54 ± 0.13} \\
\cmidrule{2-6}
 & \multirow{2}{*}{50} 
 & Text only & 88.12 ± 0.26 & 98.76 ± 0.12 & 93.07 ± 0.16  \\
 & & With 1 audio & \textbf{91.63 ± 0.21} & \textbf{99.57 ± 0.08} & \textbf{95.24 ± 0.23} \\
\midrule
\multirow{4.5}{*}{CommonVoice} & \multirow{2}{*}{1} 
& Text only & 74.96 ± 0.25& 86.01 ± 0.22 & 81.79 ± 0.15 \\
 & & With 1 audio & \textbf{76.02 ± 0.17} & \textbf{89.55 ± 0.31} & \textbf{83.56 ± 0.21} \\
\cmidrule{2-6}
 & \multirow{2}{*}{50} 
 & Text only & 76.47 ± 0.12 & 89.46 ± 0.32 & 82.33 ± 0.19 \\
 & & With 1 audio & \textbf{79.34 ± 0.23} & \textbf{91.13 ± 0.16} & \textbf{85.69 ± 0.24} \\
\bottomrule
\end{tabular}
}
\label{table:aaudio}
\end{table}

The audio optimization was performed for $n= 100$ epochs; and in each epoch, $m= 100$ Gradient Descent (GD) iterations with a learning rate of $3 \times 10^{-2}$. Four anomaly detection models, i.e., LocalOutlierFactor~\cite{cheng2019outlier}, IsolationForest~\cite{liu2008isolation}, OneClassSVM~\cite{li2003improving,khan2014one}, and AutoEncoder~\cite{chen2018autoencoder} were trained, and $N=3$ was chosen as the detection threshold. 

As shown in Table~\ref{table:baseline}, USMID consistently outperforms all baselines even with only text PII, achieving a precision of 88.12\% on LibriSpeech with 50 audio clips per person.

Additionally, USMID demonstrates notable advantages in training time and resource efficiency compared to baseline methods as shown in Table~\ref{table:costs}. It requires only 3.7 hours of training, much less than AuditMI's 80 hours, while maintaining competitive inference times. 

We also evaluate the effect of providing USMID with a real audio of the tested person. In this case, the embedding distances between the real and optimized audios of the test samples are used to perform a $2$-means clustering, adding another vote to the inference. We accordingly raise the detection threshold $N'$ to $4$. As illustrated in Table~\ref{table:aaudio}, the given audio helps to improve the performance of USMID across all tested CLAP models, showing an increase of 3.36\% on CommonVoice with 1 audio clip per person.

\subsection{Ablation Study} 

We further explore the impacts of different system parameters on the detection accuracy.

\begin{figure}[!h]
\noindent
\begin{minipage}{0.23\textwidth}
  \includegraphics[width=\linewidth]{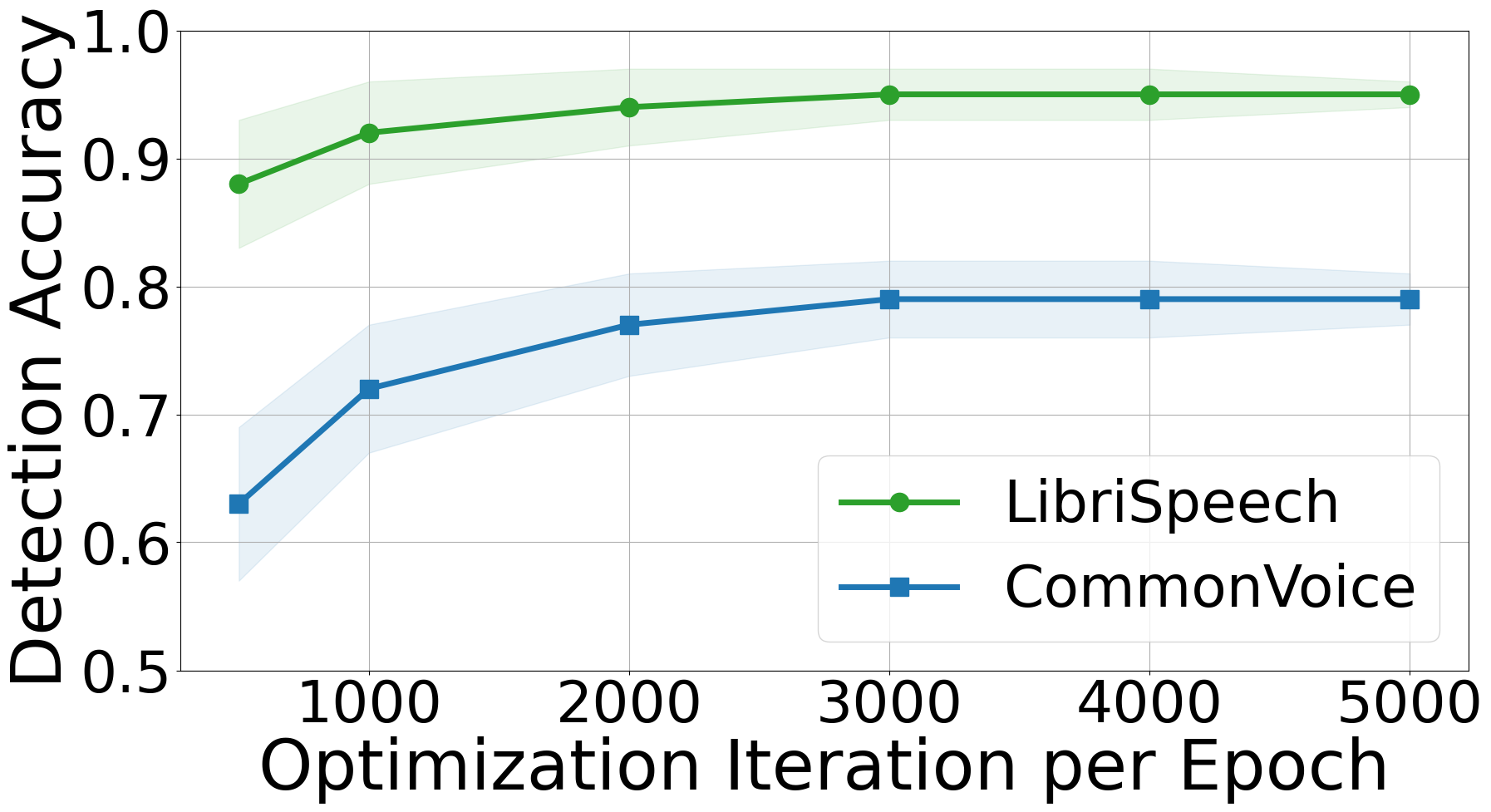}
  \caption{\small Detection accuracy for different numbers of optimization iterations per epoch.}
  \label{fig:iteration}
\end{minipage}%
\hspace*{2mm}
\begin{minipage}{0.23\textwidth}
  \includegraphics[width=\linewidth]{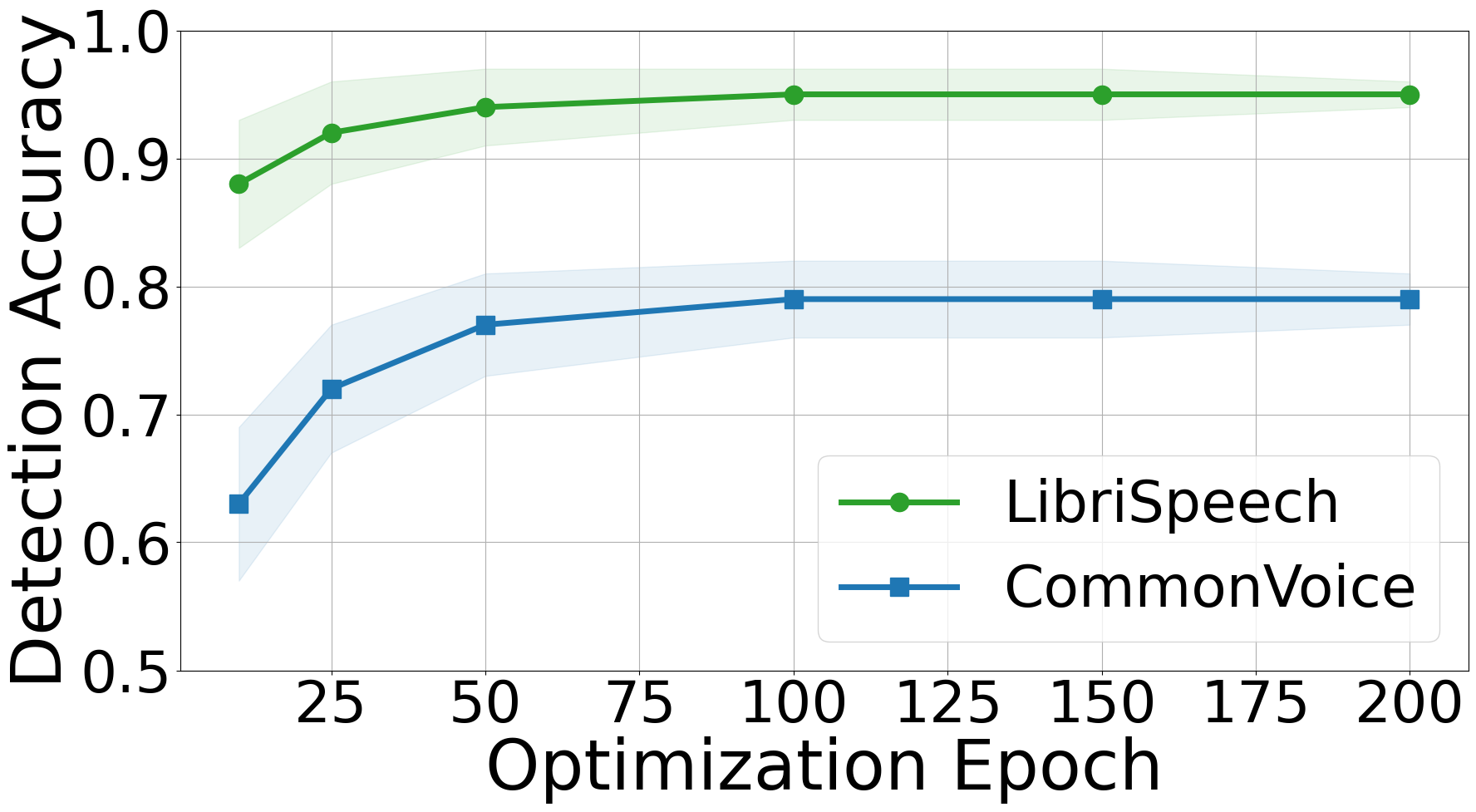}
  \caption{\small Detection accuracy for different numbers of epochs.}
  \label{fig:epoch}
\end{minipage}
\end{figure}

\begin{figure}[!h]
\noindent
\begin{minipage}{0.23\textwidth}
  \includegraphics[width=\linewidth]{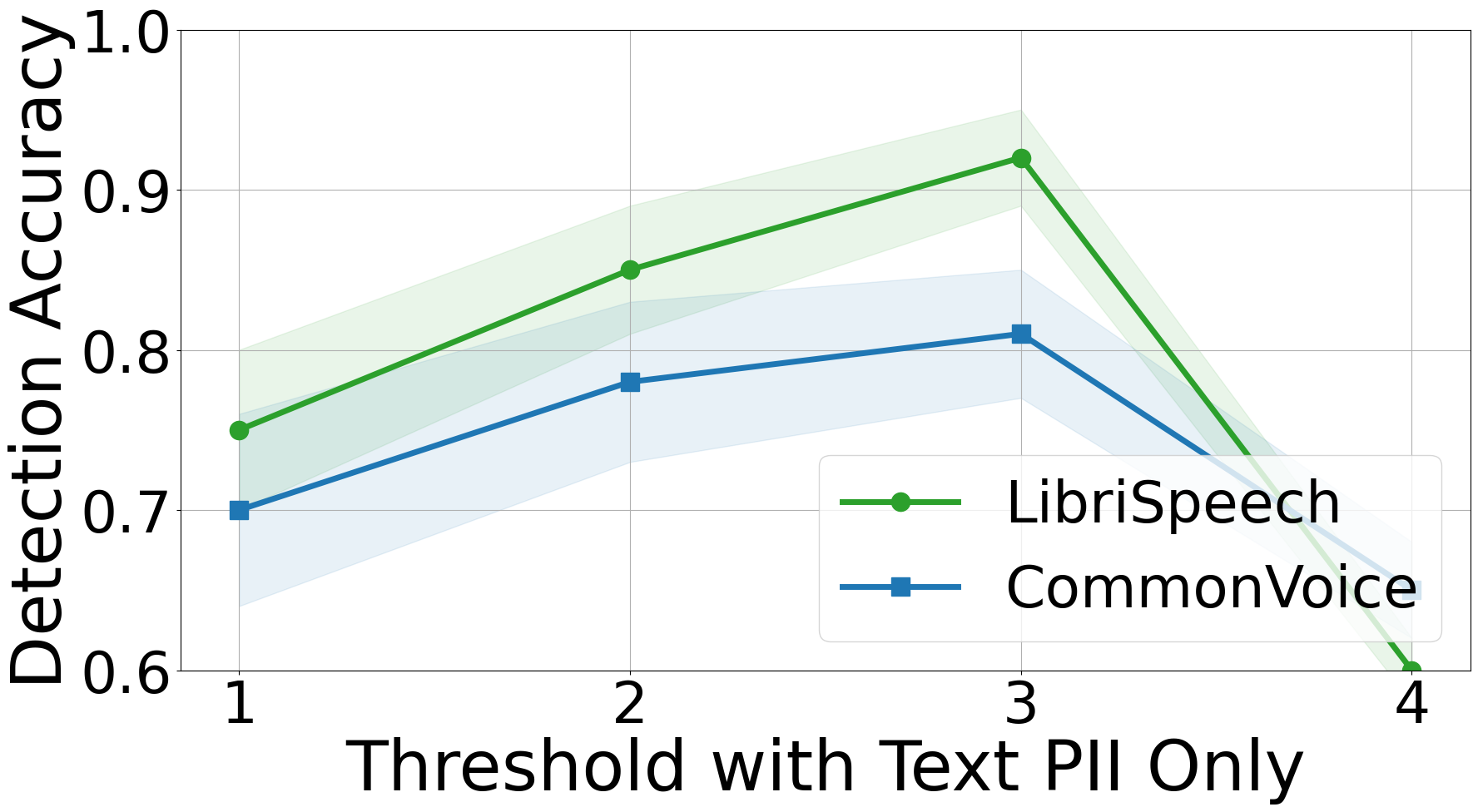}
  \caption{\small Detection accuracy with text PII only.}
\label{fig:textual}
\end{minipage}%
\hspace*{2mm}
\begin{minipage}{0.23\textwidth}
  \includegraphics[width=\linewidth]{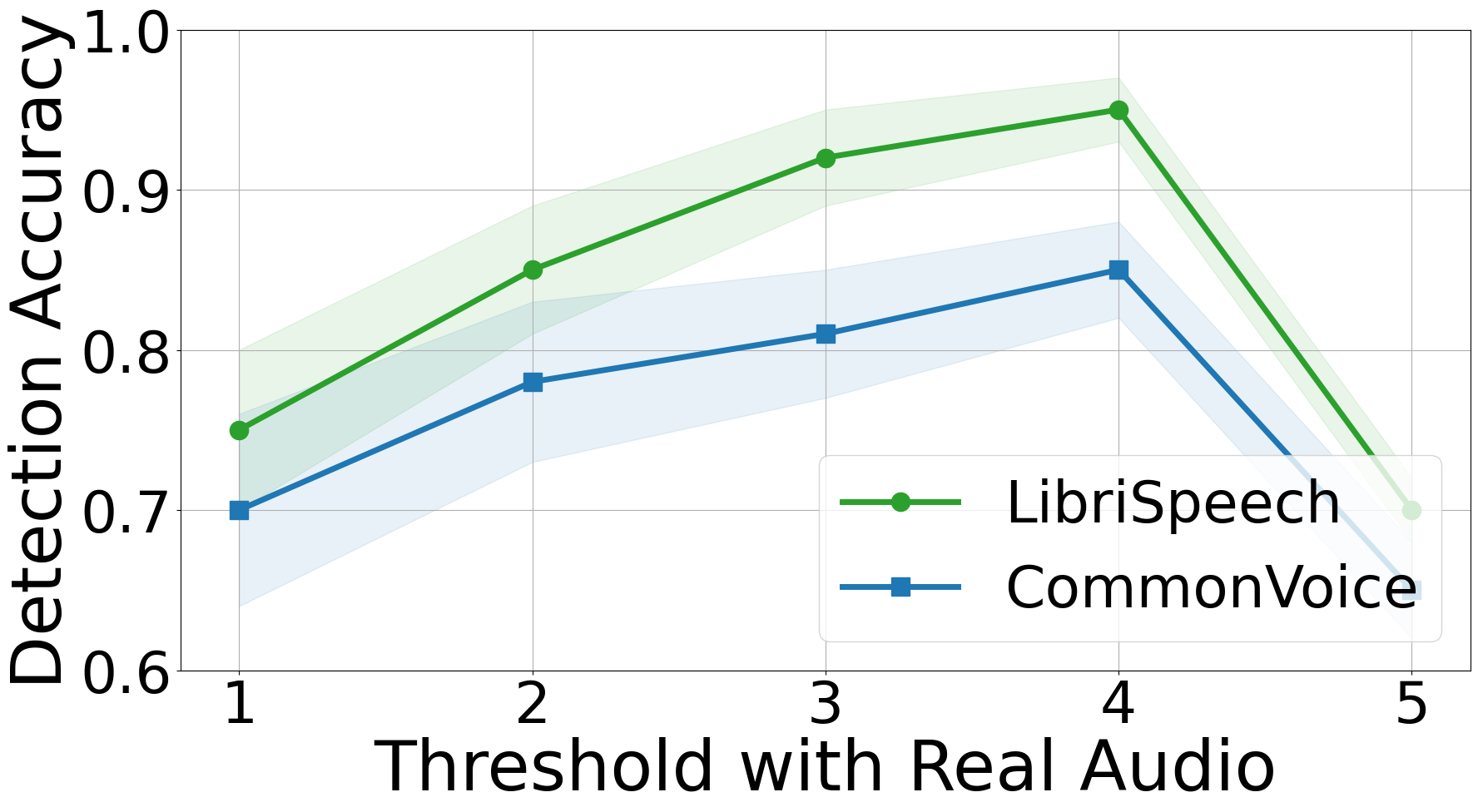}
  \caption{\small Detection accuracy with real audio for enhancement.}
   \label{fig:audio}
\end{minipage}
\end{figure}

\begin{figure}[!h]
\noindent
\begin{minipage}{0.23\textwidth}
  \includegraphics[width=\linewidth]{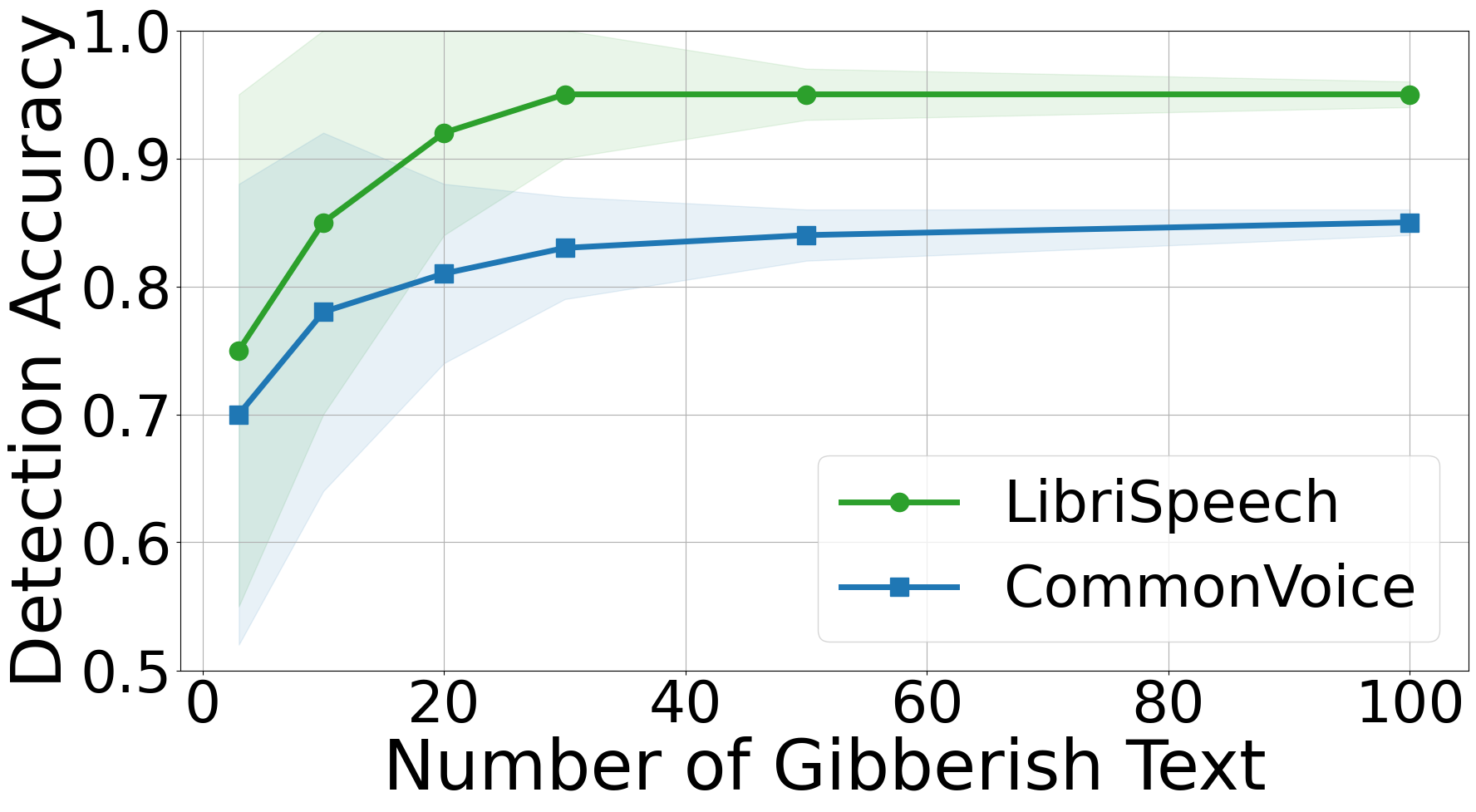}
  \caption{\small Detection accuracy for different numbers of gibberish.}
  \label{fig:gibberish}
\end{minipage}%
\hspace*{2mm}
\begin{minipage}{0.23\textwidth}
  \includegraphics[width=\linewidth]{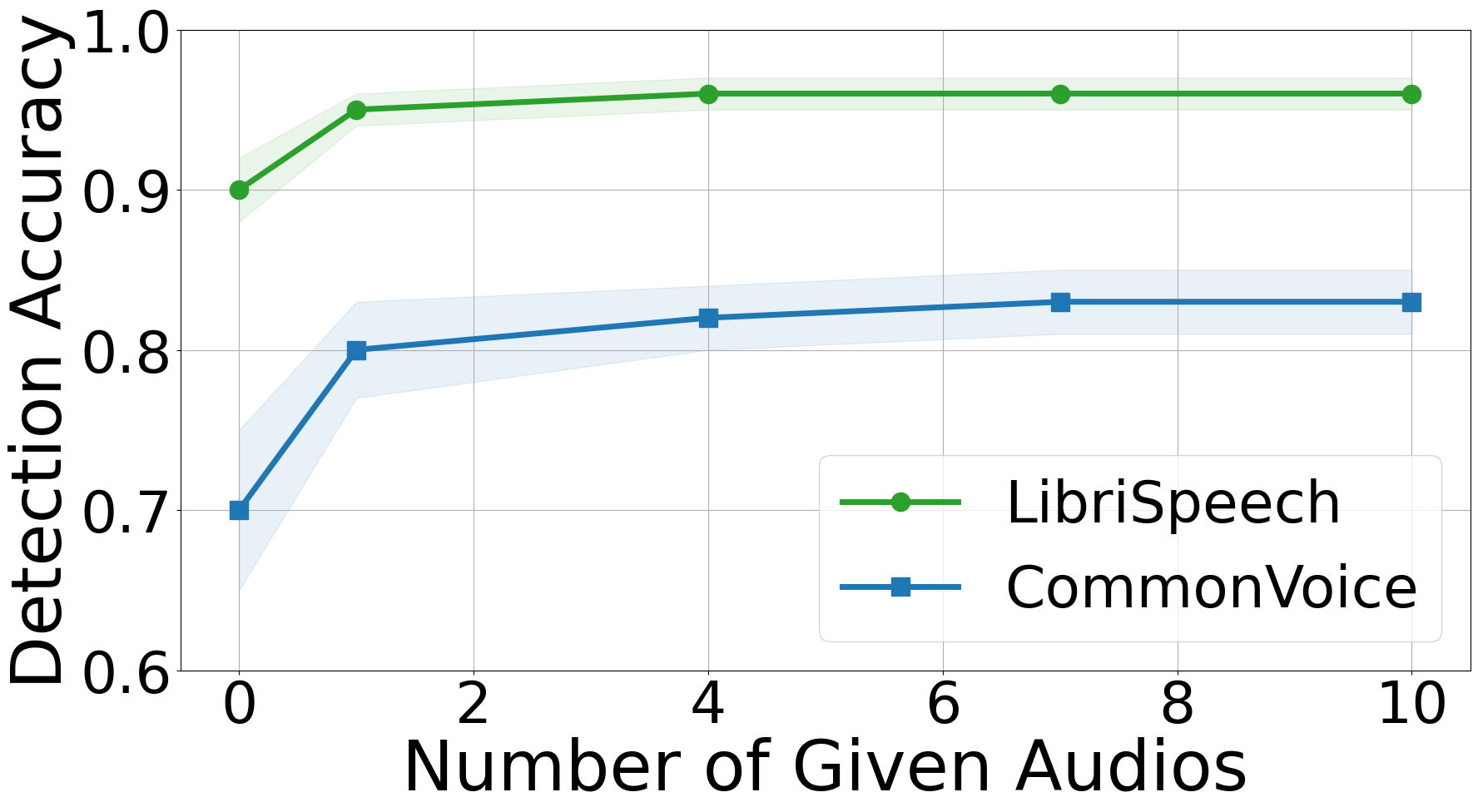}
  \caption{\small Detection accuracy for different number of real audio.}
  \label{fig:given audio}
\end{minipage}
\end{figure}

{\bf Optimization parameters.} Figure~\ref{fig:iteration} and \ref{fig:epoch} show that during feature extraction, optimizing for $n=100$ epochs, each with $m=1,000$ iterations, offers the optimal performance. Additional epochs and optimization iterations yield minimal improvements despite increased computational costs.

{\bf Detection threshold.} Figure~\ref{fig:textual} and \ref{fig:audio} show that the system achieves higher accuracy with a threshold of three votes for text-only inputs and four votes when real audio is included. A high threshold may miss anomalies, while a low threshold may incorrectly classify normal inputs as anomalies.

{\bf Number of textual gibberish.} As shown in Figure~\ref{fig:gibberish}, for different target models, the detection accuracies initially improve as the number of gibberish texts increases, and converge after using more than $50$ gibberish strings.

{\bf Number of real audios.} As shown in Figure~\ref{fig:given audio}, integrating real audios can enhance the detection accuracy; however, the improvements of using more than 1 audio are rather marginal.

\section{Defense and Covert Gibberish Generation}

In real-world scenarios, target models may target models may implement defense mechanisms to detect anomalous inputs, such as gibberish, potentially leading to misleading outputs that cause USMID to misidentify the inclusion of PII. To address this, we prompted GPT-3.5-turbo to generate fictional character backgrounds rather than mere gibberish as shown in Table~\ref{table:covertgibberish}.

\begin{table}[t]
\centering
\caption{Covert gibberish that seem to be real PII.}
\label{table:characters}
\begin{tabular}{|c|c|c|}
\hline
\textbf{Name} & \textbf{Occupation} & \textbf{Hometown} \\
\hline
Jaston Spark & Alien Biologist & Martian Oasis City \\
\hline
Carl Thunder & Climate Manipulator & Stormhaven \\
\hline
Vega Quasar & Cosmic Navigator & Starfall Galaxy \\
\hline
\end{tabular}
\label{table:covertgibberish}
\end{table}

\section{Conclusion}
\label{sec:conclusion}

This paper presents the first focused study on
membership inference detection in contrastive
language-audio pre-training models. We introduce PRMID and USMID, both of which avoid the need for computationally expensive shadow models required in traditional MIAs. Additionally, USMID is the first approach to conduct membership inference without exposing real audio samples to target CLAP models. Evaluations across various CLAP model architectures and dataset demonstrate the consistent superiority of USMID across baseline methods.

\bibliographystyle{IEEEtran}
\bibliography{reference}

\end{document}